\documentclass[10pt,a4paper,onecolumn]{article}
\usepackage{marginnote}
\usepackage{graphicx}
\usepackage{xcolor}
\usepackage{authblk,etoolbox}
\usepackage{titlesec}
\usepackage{calc}
\usepackage{tikz}
\usepackage{hyperref}
\hypersetup{colorlinks,breaklinks=true,
            urlcolor=[rgb]{0.0, 0.5, 1.0},
            linkcolor=[rgb]{0.0, 0.5, 1.0}}
\usepackage{caption}
\usepackage{tcolorbox}
\usepackage{amssymb,amsmath}
\usepackage{ifxetex,ifluatex}
\usepackage{seqsplit}
\usepackage{xstring}

\usepackage{float}
\let\origfigure\figure
\let\endorigfigure\endfigure
\renewenvironment{figure}[1][2] {
    \expandafter\origfigure\expandafter[H]
} {
    \endorigfigure
}

\usepackage{fixltx2e} 
\usepackage[
  backend=biber,
]{biblatex}
\bibliography{paper.bib}

\let\textttOrig=\texttt
\def\texttt#1{\expandafter\textttOrig{\seqsplit{#1}}}
\renewcommand{\seqinsert}{\ifmmode
  \allowbreak
  \else\penalty6000\hspace{0pt plus 0.02em}\fi}

\makeatletter
\let\href@Orig=\href
\def\href@Urllike#1#2{\href@Orig{#1}{\begingroup
    \def\Url@String{#2}\Url@FormatString
    \endgroup}}
\def\href@Notdoi#1#2{\def\tempa{#1}\def\tempb{#2}%
  \ifx\tempa\tempb\relax\href@Urllike{#1}{#2}\else
  \href@Orig{#1}{#2}\fi}
\def\href#1#2{%
  \IfBeginWith{#1}{https://doi.org}%
  {\href@Urllike{#1}{#2}}{\href@Notdoi{#1}{#2}}}
\makeatother

\usepackage[top=3.5cm, bottom=3cm, right=1.5cm, left=1.0cm,
            headheight=2.2cm, reversemp, includemp, marginparwidth=4.5cm]{geometry}

\titleformat{\section}
  {\normalfont\sffamily\Large\bfseries}
  {}{0pt}{}
\titleformat{\subsection}
  {\normalfont\sffamily\large\bfseries}
  {}{0pt}{}
\titleformat{\subsubsection}
  {\normalfont\sffamily\bfseries}
  {}{0pt}{}
\titleformat*{\paragraph}
  {\sffamily\normalsize}

\usepackage{fancyhdr}
\makeatletter
\let\ps@plain\ps@fancy
\fancyheadoffset[L]{4.5cm}
\fancyfootoffset[L]{4.5cm}

\definecolor{linky}{rgb}{0.0, 0.5, 1.0}

\newtcolorbox{repobox}
   {colback=red, colframe=red!75!black,
     boxrule=0.5pt, arc=2pt, left=6pt, right=6pt, top=3pt, bottom=3pt}

\newcommand{\ExternalLink}{%
   \tikz[x=1.2ex, y=1.2ex, baseline=-0.05ex]{%
       \begin{scope}[x=1ex, y=1ex]
           \clip (-0.1,-0.1)
               --++ (-0, 1.2)
               --++ (0.6, 0)
               --++ (0, -0.6)
               --++ (0.6, 0)
               --++ (0, -1);
           \path[draw,
               line width = 0.5,
               rounded corners=0.5]
               (0,0) rectangle (1,1);
       \end{scope}
       \path[draw, line width = 0.5] (0.5, 0.5)
           -- (1, 1);
       \path[draw, line width = 0.5] (0.6, 1)
           -- (1, 1) -- (1, 0.6);
       }
   }

\patchcmd{\@maketitle}{center}{flushleft}{}{}
\patchcmd{\@maketitle}{center}{flushleft}{}{}
\patchcmd{\@maketitle}{\LARGE}{\LARGE\sffamily}{}{}
\def\maketitle{{%
  
  \AB@maketitle}}
\makeatletter
\renewcommand\AB@affilsepx{ \protect\Affilfont}
\renewcommand\AB@affilnote[1]{{\bfseries #1}\hspace{3pt}}
\renewcommand{\affil}[2][]%
   {\newaffiltrue\let\AB@blk@and\AB@pand
      \if\relax#1\relax\def\AB@note{\AB@thenote}\else\def\AB@note{#1}%
        \setcounter{Maxaffil}{0}\fi
        \begingroup
        \let\href=\href@Orig
        \let\texttt=\textttOrig
        \let\protect\@unexpandable@protect
        \def\thanks{\protect\thanks}\def\footnote{\protect\footnote}%
        \@temptokena=\expandafter{\AB@authors}%
        {\def\\{\protect\\\protect\Affilfont}\xdef\AB@temp{#2}}%
         \xdef\AB@authors{\the\@temptokena\AB@las\AB@au@str
         \protect\\[\affilsep]\protect\Affilfont\AB@temp}%
         \gdef\AB@las{}\gdef\AB@au@str{}%
        {\def\\{, \ignorespaces}\xdef\AB@temp{#2}}%
        \@temptokena=\expandafter{\AB@affillist}%
        \xdef\AB@affillist{\the\@temptokena \AB@affilsep
          \AB@affilnote{\AB@note}\protect\Affilfont\AB@temp}%
      \endgroup
       \let\AB@affilsep\AB@affilsepx
}
\makeatother

\renewcommand\Affilfont{\sffamily\small\mdseries}
\ifnum 0\ifxetex 1\fi\ifluatex 1\fi=0 
  \usepackage[T1]{fontenc}
  \usepackage[utf8]{inputenc}

\else 
  \ifxetex
    \usepackage{mathspec}
    \usepackage{fontspec}

  \else
    \usepackage{fontspec}
  \fi
  \defaultfontfeatures{Ligatures=TeX,Scale=MatchLowercase}

\fi
\IfFileExists{upquote.sty}{\usepackage{upquote}}{}
\IfFileExists{microtype.sty}{%
\usepackage{microtype}
\UseMicrotypeSet[protrusion]{basicmath} 
}{}

\usepackage{hyperref}
\hypersetup{unicode=true,
            pdftitle={ParaMonte: A high-performance serial/parallel Monte Carlo simulation library for C, C++, Fortran},
            pdfborder={0 0 0},
            breaklinks=true}
\let\addcontentslineOrig=\addcontentsline
\def\addcontentsline#1#2#3{\bgroup
  \let\texttt=\textttOrig\addcontentslineOrig{#1}{#2}{#3}\egroup}
\let\markbothOrig\markboth
\def\markboth#1#2{\bgroup
  \let\texttt=\textttOrig\markbothOrig{#1}{#2}\egroup}
\let\markrightOrig\markright
\def\markright#1{\bgroup
  \let\texttt=\textttOrig\markrightOrig{#1}\egroup}

\usepackage{graphicx,grffile}
\makeatletter
\def\maxwidth{\ifdim\Gin@nat@width>\linewidth\linewidth\else\Gin@nat@width\fi}
\def\maxheight{\ifdim\Gin@nat@height>\textheight\textheight\else\Gin@nat@height\fi}
\makeatother
\setkeys{Gin}{width=\maxwidth,height=\maxheight,keepaspectratio}
\IfFileExists{parskip.sty}{%
\usepackage{parskip}
}{
\setlength{\parindent}{0pt}
\setlength{\parskip}{6pt plus 2pt minus 1pt}
}
\providecommand{\tightlist}{%
  \setlength{\itemsep}{0pt}\setlength{\parskip}{0pt}}
\ifx\paragraph\undefined\else
\let\oldparagraph\paragraph
\renewcommand{\paragraph}[1]{\oldparagraph{#1}\mbox{}}
\fi
\ifx\subparagraph\undefined\else
\let\oldsubparagraph\subparagraph
\renewcommand{\subparagraph}[1]{\oldsubparagraph{#1}\mbox{}}
\fi

\title{ParaMonte: A high-performance serial/parallel Monte Carlo simulation
library for C, C++, Fortran}

        \author[1, 2]{Amir Shahmoradi}
          \author[1]{Fatemeh Bagheri}
    
      \affil[1]{Department of Physics, The University of Texas, Arlington, TX}
      \affil[2]{Data Science Program, The University of Texas, Arlington, TX}
  \date{\vspace{-7ex}}

\begin{document}
\maketitle

\marginpar{

  \begin{flushleft}
  \sffamily\small

  {\bfseries DOI:} \href{https://doi.org/}{\color{linky}{}}

  \vspace{2mm}

  {\bfseries Software}
  \begin{itemize}
    \setlength\itemsep{0em}
    \item \href{}{\color{linky}{Review}} \ExternalLink
    \item \href{https://github.com/cdslaborg/paramonte}{\color{linky}{Repository}} \ExternalLink
    \item \href{}{\color{linky}{Archive}} \ExternalLink
  \end{itemize}

  \vspace{2mm}

  \par\noindent\hrulefill\par

  \vspace{2mm}

  {\bfseries Editor:} \href{}{} \ExternalLink \\
  \vspace{1mm}
    \vspace{2mm}

  {\bfseries Submitted:} 29 September 2020\\
  {\bfseries Published:} 

  \vspace{2mm}
  {\bfseries License}\\
  Authors of papers retain copyright and release the work under a Creative Commons Attribution 4.0 International License (\href{http://creativecommons.org/licenses/by/4.0/}{\color{linky}{CC BY 4.0}}).

  \end{flushleft}
}

\section{Summary}\label{summary}

ParaMonte (standing for Parallel Monte Carlo) is a serial and
MPI/Coarray-parallelized library of Monte Carlo routines for sampling
mathematical objective functions of arbitrary-dimensions, in particular,
the posterior distributions of Bayesian models in data science, Machine
Learning, and scientific inference. The ParaMonte library has been
developed with the design goal of unifying the \textbf{automation},
\textbf{accessibility}, \textbf{high-performance}, \textbf{scalability},
and \textbf{reproducibility} of Monte Carlo simulations. The current
implementation of the library includes \textbf{ParaDRAM}, a
\textbf{Para}llel \textbf{D}elyaed-\textbf{R}ejection \textbf{A}daptive
\textbf{M}etropolis Markov Chain Monte Carlo sampler, accessible from a
wide range of programming languages including C, C++, Fortran, with a
unified Application Programming Interface and simulation environment
across all supported programming languages. The ParaMonte library is
MIT-licensed and is permanently located and maintained at
\url{https://github.com/cdslaborg/paramonte}.

\section{Statement of need}\label{statement-of-need}

Monte Carlo simulation techniques, in particular, the Markov Chain Monte
Carlo (MCMC) are among the most popular methods of quantifying
uncertainty in scientific inference problems. Extensive work has been
done over the past decades to develop Monte Carlo simulation programming
environments that aim to partially or fully automate the problem of
uncertainty quantification via Markov Chain Monte Carlo simulations.
Example open-source libraries in C/C++/Fortran include \texttt{MCSim} in
C (Bois, 2009), \texttt{MCMCLib} and \texttt{QUESO} (Prudencio \&
Schulz, 2012) libraries in C++, and \texttt{mcmcf90} in Fortran (Haario,
Laine, Mira, \& Saksman, 2006). These packages, however, mostly serve
the users of one particular programming language environment. Some are
able to perform only serial simulations while others are inherently
parallelized. Furthermore, majority of the existing packages have
significant dependencies on other external libraries. Such dependencies
can potentially make the build process of the packages an extremely
complex and arduous task due to software version incompatibilities, a
phenomenon that has become known as the \emph{dependency-hell} among
software developers.

The ParaMonte library presented in this work aims to address the
aforementioned problems by providing a standalone high-performance
serial/parallel Monte Carlo simulation environment with the following
principal design goals,

\begin{itemize}
\tightlist
\item
  \textbf{Full automation} of the library's build process and all Monte
  Carlo simulations to ensure the highest level of user-friendliness of
  the library and minimal time investment requirements for building,
  running, and post-processing of the Monte Carlo and MCMC
  simulations.\\
\item
  \textbf{Interoperability} of the core of the library with as many
  programming languages as currently possible, including C, C++,
  Fortran, as well as MATLAB and Python via the
  \texttt{ParaMonte::MATLAB} and \texttt{ParaMonte::Python} libraries.\\
\item
  \textbf{High-Performance}, meticulously-low-level, implementation of
  the library that \textbf{guarantees the fastest-possible Monte Carlo
  simulations}, without compromising the reproducibility of the
  simulation or the extensive external reporting of the simulation
  progress and results.\\
\item
  \textbf{Parallelizability} of all simulations via both MPI and
  PGAS/Coarray communication paradigms while \textbf{requiring
  zero-parallel-coding efforts from the user}.\\
\item
  \textbf{Zero external-library dependencies} to ensure hassle-free
  library builds and Monte Carlo simulation runs.\\
\item
  \textbf{Fully-deterministic reproducibility} and
  \textbf{automatically-enabled restart functionality} for all ParaMonte
  simulations, up to 16 digits of decimal precision if requested by the
  user.\\
\item
  \textbf{Comprehensive-reporting and post-processing} of each
  simulation and its results, as well as their efficient compact storage
  in external files to ensure the simulation results will be
  comprehensible and reproducible at any time in the distant future.
\end{itemize}

\section{The Build process}\label{the-build-process}

The ParaMonte library is permanently located on GitHub and is available
to view at: \url{https://github.com/cdslaborg/paramonte}. The build
process of the library is fully automated. Extensive detailed
instructions are also available on the
\href{https://www.cdslab.org/paramonte/}{documentation website of the
library}.

For the convenience of the users, each release of the ParaMonte
library's source code also includes prebuilt, ready-to-use, copies of
the library for \texttt{x64} architecture on Windows, Linux, macOS, in
all supported programming languages, including C, C++, Fortran. These
prebuilt libraries automatically ship with language-specific example
codes and build scripts that fully automate the process of building and
running the examples.

Where the prebuilt libraries cannot be used, the users can simply call
the Bash and Batch build-scripts that are provided in the source code of
the library to fully automate the build process of the library. The
ParaMonte build scripts are capable of automatically installing any
missing components that may be required for the library's successful
build, including the GNU C/C++/Fortran compilers and the \texttt{cmake}
build software, as well as the MPI/Coarray parallelism libraries. All of
these tasks are performed with the explicit permission granted by the
user. The ParaMonte build scripts are heavily inspired by the impressive
\texttt{OpenCoarrays} software (Fanfarillo et al., 2014) developed and
maintained by the \href{http://www.sourceryinstitute.org/}{Sourcery
Institute}.

\section{The ParaDRAM sampler}\label{the-paradram-sampler}

The current implementation of the ParaMonte library includes the
\textbf{Para}llel \textbf{D}elayed-\textbf{R}ejection \textbf{A}daptive
\textbf{M}etropolis Markov Chain Monte Carlo
(\textbf{\texttt{ParaDRAM}}) sampler (Shahmoradi \& Bagheri, 2020),
(Shahmoradi \& Bagheri, 2020a), (Shahmoradi \& Bagheri, 2020b),
(Kumbhare \& Shahmoradi, 2020), and several other samplers whose
development is in progress as of writing this manuscript. The ParaDRAM
algorithm is a variant of the DRAM algorithm of (Haario et al., 2006)
and can be used in serial or parallel mode.

In brief, the ParaDRAM sampler continuously adapts the shape and scale
of the proposal distribution throughout the simulation to increase the
efficiency of the sampler. This is in contrast to the traditional MCMC
samplers where the proposal distribution remains fixed throughout the
simulation. The ParaDRAM sampler provides a highly customizable MCMC
simulation environment whose complete description goes beyond the scope
and limits of this manuscript. All of these simulation specifications
are, however, expensively explained and discussed on
\href{https://www.cdslab.org/paramonte/}{the documentation website of
the ParaMonte library}. The description of all of these specifications
are also automatically provided in the output \texttt{*\_report.txt}
files of every simulation performed by the ParaMonte samplers.

\subsection{Monitoring Convergence}\label{monitoring-convergence}

Although the continuous adaptation of the proposal distribution
increases the sampling efficiency of the ParaDRAM sampler, it breaks the
ergodicity and reversibility conditions of the Markov Chain Monte Carlo
methods. Nevertheless, the convergence of the resulting pseudo-Markov
chain to the target density function is guaranteed as long as the amount
of adaptation of the proposal distribution decreases monotonically
throughout the simulation (Shahmoradi \& Bagheri, 2020).

Ideally, the diminishing adaptation criterion of the adaptive MCMC
methods can be monitored by measuring the \emph{total variation
distance} (TVD) between subsequent adaptively-updated proposal
distributions. Except for trivial cases, however, the analytical or
numerical computation of TVD is almost always intractable. To circumvent
this problem, we have introduced a novel technique in the ParaDRAM
algorithm to continuously measure the amount of adaptation of the
proposal distribution throughout adaptive MCMC simulations. This is done
by computing an upper bound on the value of TVD instead of a direct
computation of the TVD.

The mathematics of computing this \texttt{AdaptationMeasure} upper bound
is extensively detailed in (Shahmoradi \& Bagheri, 2020), (Shahmoradi \&
Bagheri, 2020a). The computed upper bound is always a real number
between 0 and 1, with 0 indicating the identity of two proposal
distributions and 1 indicating completely different proposal
distributions. The \texttt{AdaptationMeasure} is automatically computed
with every proposal adaptation and is reported to the output chain files
for all ParaDRAM simulations. It can be subsequently visualized to
ensure the diminishing adaptation criterion of the ParaDRAM sampler.

\begin{figure}
\centering
\includegraphics{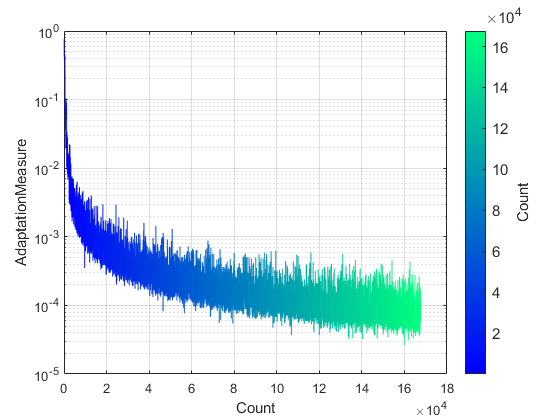}
\caption{An illustration of the diminishing adaptation of the proposal
distribution of the ParaDRAM sampler for an example problem of sampling
a 4-dimensional Multivariate Normal distribution. The
monotonically-decreasing adaptivity evidenced in this plot guarantees
the Markovian property and the asymptotic ergodicity of the resulting
Markov chain from the ParaDRAM sampler.\label{fig:adaptationMeasure}}
\end{figure}

\begin{figure}
\centering
\includegraphics{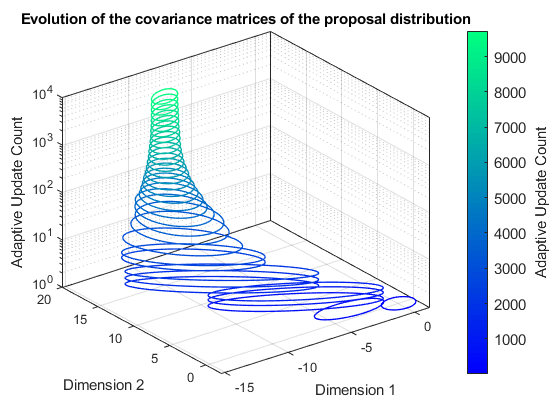}
\caption{A 3-dimensional illustration of the dynamic adaptation of the
covariance matrix of the 4-dimensional MultiVariate Normal (MVN)
proposal distribution of the ParaDRAM sampler for an example problem of
sampling a 4-dimensional MVN distribution.\label{fig:covmatEvolution}}
\end{figure}

Figure \autoref{fig:adaptationMeasure} depicts the evolution of the
adaptation measure for an example problem of sampling a 4-dimensional
MultiVariate Normal Distribution. A non-diminishing evolution of the
\texttt{AdaptationMeasure} can be also a strong indicator of the lack of
convergence the Markov chain to the target density. The evolution of the
covariance matrices of the proposal distribution for the same sampling
problem is shown in Figure \autoref{fig:covmatEvolution}.

\subsection{Parallelism}\label{parallelism}

Two modes of parallelism are currently implemented for all ParaMonte
samplers,

\begin{itemize}
\item
  The \textbf{Perfect Parallelism} (multi-Chain): In this mode,
  independent instances of the adaptive MCMC sampler run concurrently.
  Once all simulations are complete, the ParaDRAM sampler compares the
  output samples from all processors with each other to ensure no
  evidence for a lack of convergence to the target density exists in any
  of the output chains.
\item
  The \textbf{Fork-Join Parallelism} (single-Chain): In this mode, a
  single processor is responsible for collecting and dispatching
  information, generated by all processors, to create a single Markov
  Chain of all visited states with the help of all processors.
\end{itemize}

For each parallel simulation in the Fork-Join mode, the ParaMonte
samplers automatically compute the speedup gained compared to the serial
mode. In addition, the speedup for a wide range of the number of
processors is also automatically computed and reported in the output
\texttt{*\_report.txt} files that are automatically generated for all
simulations. The processor contributions to the construction of each
chain are also reported along with output visited states in the output
\texttt{*\_chain.*} files. These reports are particularly useful for
finding the optimal number of processors for a given problem at hand, by
first running a short simulation to predict the optimal number of
processors from the sampler's output information, followed by the
production run using the optimal number of processors. For a
comprehensive description and algorithmic details see (Shahmoradi \&
Bagheri, 2020), (Shahmoradi \& Bagheri, 2020a).

\begin{figure}
\centering
\includegraphics{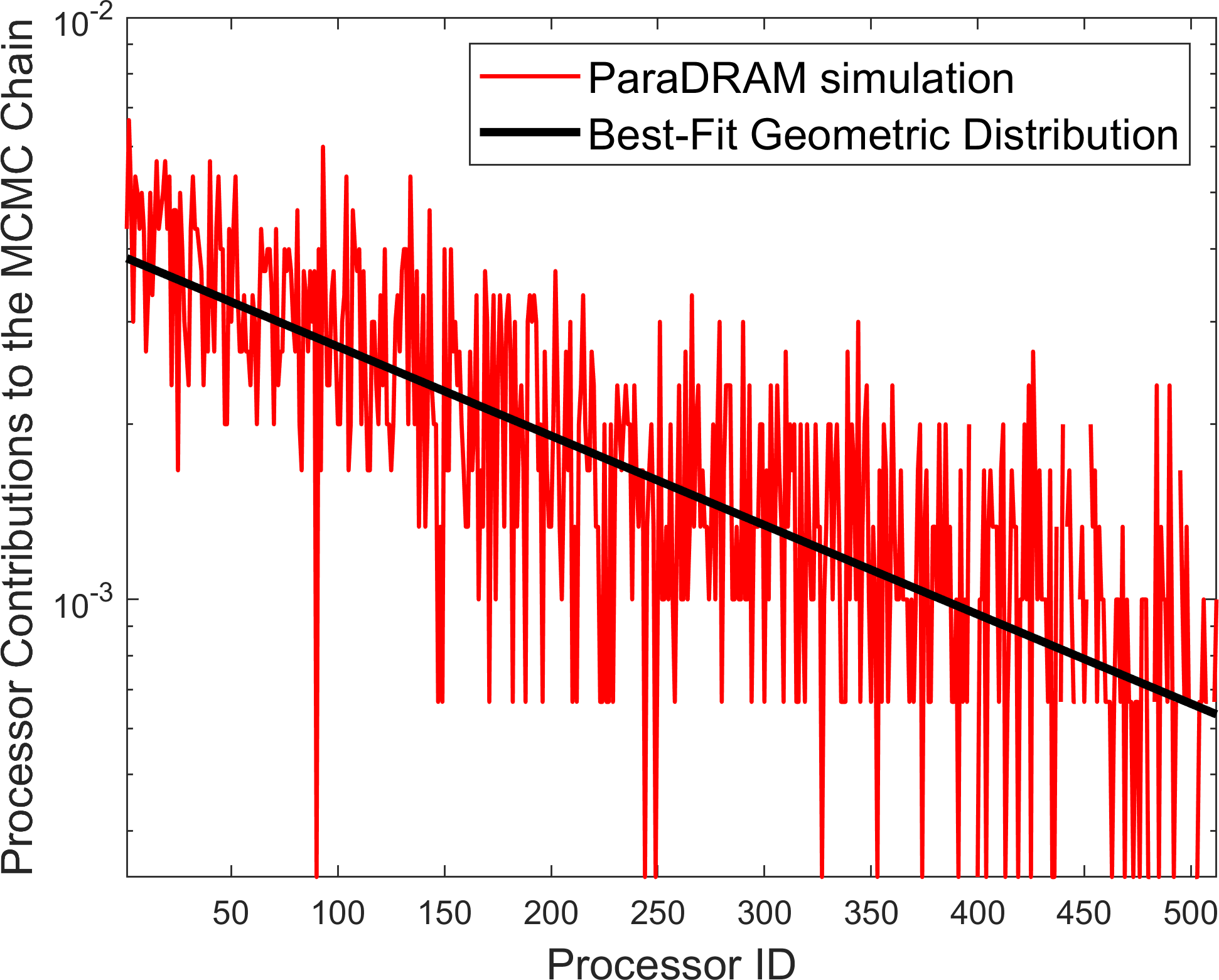}
\caption{An illustration of the contributions of 512 Intel Xeon Phi 7250
processors to a ParaMonte-ParaDRAM simulation parallelized via the
Fork-Join paradigm. The predicted best-fit Geometric distribution from
the post-processing phase of the ParaDRAM simulation is shown by the
black line. The data used in this figure is automatically generated for
each parallel simulation performed via any of the ParaMonte samplers.
\label{fig:procContribution512}}
\end{figure}

\begin{figure}
\centering
\includegraphics{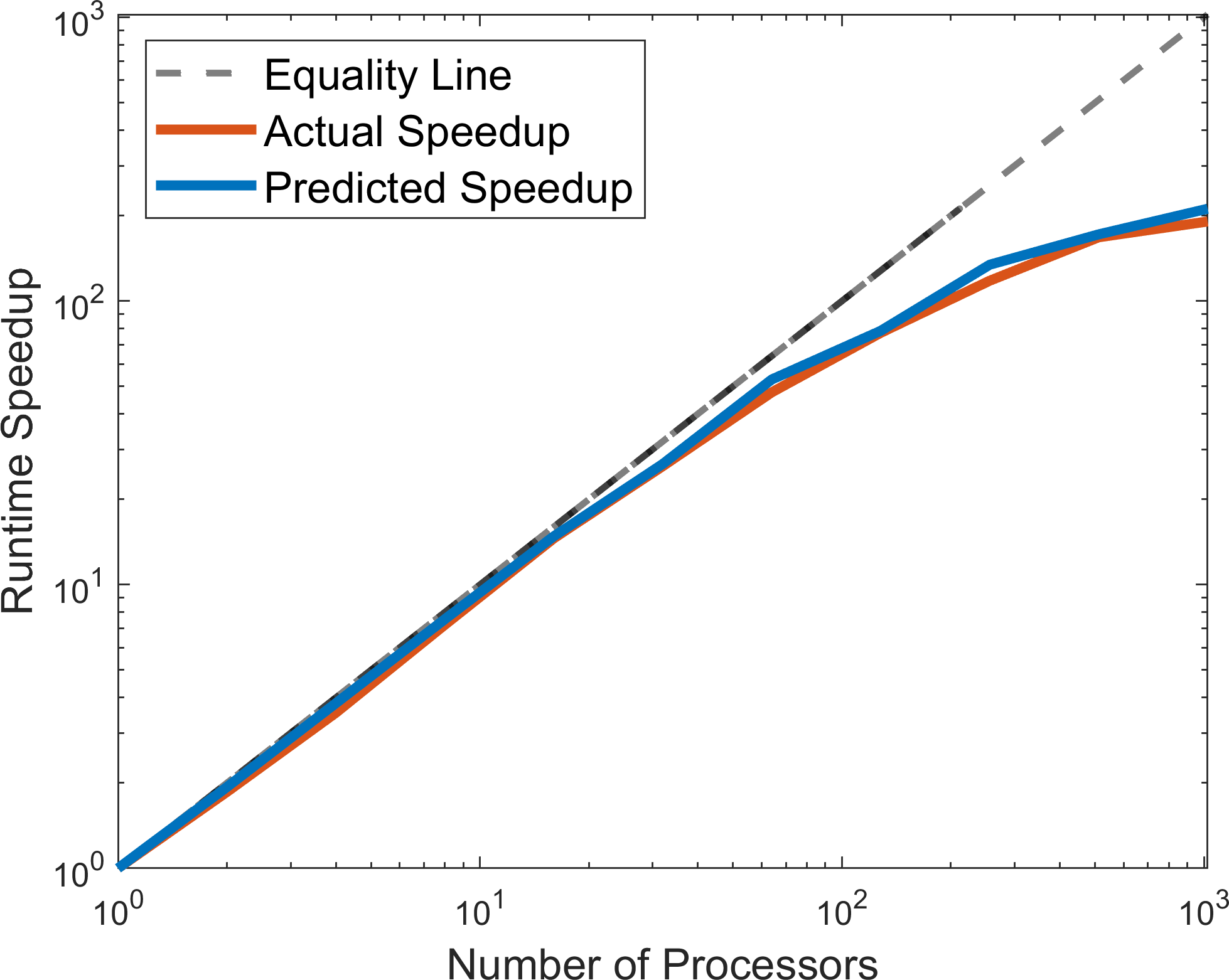}
\caption{A comparison of the actual strong scaling behavior of an
example ParaMonte-ParaDRAM simulation from 1 to 1088 processors with the
strong-scaling behavior predicted during the post-processing phases of
ParaDRAM simulations. The data used in this figure is automatically
generated for each parallel simulation performed via any of the
ParaMonte samplers.\label{fig:PredictedSpeedupActualSpeedup}}
\end{figure}

As we argue in (Shahmoradi \& Bagheri, 2020, Shahmoradi \& Bagheri
(2020a)), the contribution of the processors to the construction of a
Markov Chain in the Fork-Join parallelism paradigm follows a Geometric
distribution. Figure \autoref{fig:procContribution512} depicts the
processor contributions to an example ParaDRAM simulation of a variant
of Himmelblau's function and the Geometric fit to the distribution of
processor contributions. Figure
\autoref{fig:PredictedSpeedupActualSpeedup} illustrates an example
predicted strong-scaling behavior of the sampler and the predicted
speedup by the sampler for a range of processor counts.

\subsection{Efficient compact storage of the output
chain}\label{efficient-compact-storage-of-the-output-chain}

Efficient continuous external storage of the output of ParaDRAM
simulations is essential for both the post-processing of the results and
the restart functionality of the simulations, should any interruptions
happen at runtime. However, as the number of dimensions or the
complexity of the target density increases, such external storage of the
output can easily become a challenge and a bottleneck in the speed of an
otherwise high-performance ParaDRAM sampler. Given the
currently-available computational technologies, input/ouput (IO) to
external hard-drives can be 2-3 orders of magnitude slower than the
Random Access Memory (RAM) storage.

To alleviate the effects of such external-IO speed bottlenecks, the
ParaDRAM sampler implements a novel method of carefully storing the
resulting MCMC chains in a small \emph{compact}, yet ASCII
human-readable, format in external output files. This
\textbf{compact-chain} (as opposed to the \textbf{verbose
(Markov)-chain}) format leads to significant speedup of the simulation
while requiring 4-100 times less external memory to store the chains in
the external output files. The exact amount of reduction in the external
memory usage depends on the efficiency of the sampler. Additionally, the
format of output file can be set by the user to \texttt{binary}, further
reducing the memory foot-print of the simulation while increasing the
simulation speed. The implementation details of this compact-chain
format are extensively discussed in (Shahmoradi \& Bagheri, 2020),
(Shahmoradi \& Bagheri, 2020a).

\subsection{Sample refinement}\label{sample-refinement}

In addition to the output \texttt{*\_progress.txt},
\texttt{*\_report.txt}, \texttt{*\_chain.txt} files, each ParaDRAM
sampling generates a \texttt{*\_sample.txt} file containing the final
refined decorrelated sample from the objective function. To do so, the
sampler computes the Integrated AutoCorrelation (IAC) of the chain along
each dimension of the domain of the objective function. However, the
majority of existing methods for the calculation of IAC tend to
underestimate this quantity.

Therefore, to ensure the final sample resulting from a ParaDRAM
simulation is fully decorrelated, we have implemented a novel approach
that aggressively and recursively refines the resulting Markov chain
from a ParaDRAM simulation until no trace of autocorrelation is left in
the final refined sample. This approach optionally involves two separate
phases of sample refinement,

\begin{enumerate}
\def\labelenumi{\arabic{enumi}.}
\tightlist
\item
  At the first stage, the Markov chain is decorrelated recursively, for
  as long as needed, based on the IAC of its compact format, where only
  the the uniquely-visited states are kept in the (compact) chain.\\
\item
  Once the Markov chain is refined such that its compact format is fully
  decorrelated, the second phase of decorrelation begins, during which
  the Markov chain is decorrelated based on the IAC of the chain in its
  verbose (Markov) format. This process is repeated recursively for as
  long as there is any residual autocorrelation in the refined sample.
\end{enumerate}

We have empirically noticed, via numerous experimentations, that this
recursive aggressive approach is superior to other existing methods of
sample refinement in generating final refined samples that are neither
too small in size nor autocorrelated.

\subsection{The restart functionality}\label{the-restart-functionality}

Each ParaMonte sampler is automatically capable of restarting an
existing interrupted simulation, whether in serial or parallel. All that
is required is to rerun the interrupted simulation with the same output
file names. The ParaMonte samplers automatically detect the presence of
an incomplete simulation in the output files and restart the simulation
from where it was left off. Furthermore, if the user sets the seed of
the random number generator of sampler prior to running the simulation,
\textbf{the ParaMonte samplers are capable of regenerating the same
chain that would have been produced if the simulation had not been
interrupted in the first place}. Such \emph{fully-deterministic
reproducibility into-the-future} is guaranteed with 16 digits of decimal
precision for the results of any ParaMonte simulation. To our knowledge,
this is a unique feature of the ParaMonte library that does not appear
to exist in any of the contemporary libraries for Markov Chain Monte
Carlo simulations.

\section{Documentation and
Repository}\label{documentation-and-repository}

The ParaMonte library was originally developed in 2012 and remained in
private use for the research needs of the developers of the library
(Shahmoradi, 2013), (Shahmoradi, 2013), (Shahmoradi \& Nemiroff, 2014),
(Shahmoradi \& Nemiroff, 2015), (Shahmoradi \& Nemiroff, 2019),
(Osborne, Shahmoradi, \& Nemiroff, 2020), (Osborne, Shahmoradi, \&
Nemiroff, 2020). The library was further developed in 2018 and made
available to the public with its first official release in 2020.

Extensive documentation and examples in C, C++, Fortran (as well as
other programming languages) are available on the documentation website
of the library at: \url{https://www.cdslab.org/paramonte/}. The
ParaMonte library is MIT-licensed and is permanently located and
maintained at \url{https://github.com/cdslaborg/paramonte}.

\section{Acknowledgements}\label{acknowledgements}

We thank the Texas Advanced Computing Center for providing the
supercomputer time for testing and development of this library.

\section*{References}\label{references}
\addcontentsline{toc}{section}{References}

\hypertarget{refs}{}
\hypertarget{ref-Bois:2009}{}
Bois, F. Y. (2009). GNU mcsim: Bayesian statistical inference for
sbml-coded systems biology models. \emph{Bioinformatics}, \emph{25}(11),
1453--1454.

\hypertarget{ref-Fanfarillo:2014}{}
Fanfarillo, A., Burnus, T., Cardellini, V., Filippone, S., Nagle, D., \&
Rouson, D. (2014). OpenCoarrays: Open-source transport layers supporting
coarray fortran compilers. In \emph{Proceedings of the 8th international
conference on partitioned global address space programming models} (pp.
1--11).

\hypertarget{ref-Haario:2006}{}
Haario, H., Laine, M., Mira, A., \& Saksman, E. (2006). DRAM: Efficient
adaptive mcmc. \emph{Statistics and computing}, \emph{16}(4), 339--354.

\hypertarget{ref-Kumbhare:2020}{}
Kumbhare, S., \& Shahmoradi, A. (2020). Parallel adapative monte carlo
optimization, sampling, and integration in c/c++, fortran, matlab, and
python. \emph{Bulletin of the American Physical Society}.

\hypertarget{ref-OsborneARXIV:2020}{}
Osborne, J. A., Shahmoradi, A., \& Nemiroff, R. J. (2020). A multilevel
empirical bayesian approach to estimating the unknown redshifts of 1366
batse catalog long-duration gamma-ray bursts.

\hypertarget{ref-OsborneADS:2020}{}
Osborne, J. A., Shahmoradi, A., \& Nemiroff, R. J. (2020). A Multilevel
Empirical Bayesian Approach to Estimating the Unknown Redshifts of 1366
BATSE Catalog Long-Duration Gamma-Ray Bursts. \emph{arXiv e-prints},
arXiv:2006.01157.

\hypertarget{ref-Alexander:2012}{}
Prudencio, E., \& Schulz, K. (2012). The parallel C++ statistical
library queso: Quantification of uncertainty for estimation, simulation
and optimization. In M. Alexander, P. D'Ambra, A. Belloum, G. Bosilca,
M. Cannataro, M. Danelutto, B. Martino, et al. (Eds.), \emph{Euro-par
2011: Parallel processing workshops}, Lecture notes in computer science
(Vol. 7155, pp. 398--407). Springer Berlin Heidelberg.
ISBN:~978-3-642-29736-6

\hypertarget{ref-Shahmoradi:2013}{}
Shahmoradi, A. (2013). A multivariate fit luminosity function and world
model for long gamma-ray bursts. \emph{The Astrophysical Journal},
\emph{766}(2), 111.

\hypertarget{ref-ShahmoradiA:2013}{}
Shahmoradi, A. (2013). Gamma-Ray bursts: Energetics and Prompt
Correlations. \emph{arXiv e-prints}, arXiv:1308.1097.

\hypertarget{ref-ShahmoradiGS:2020}{}
Shahmoradi, A., \& Bagheri, F. (2020). ParaDRAM: A cross-language
toolbox for parallel high-performance delayed-rejection adaptive
metropolis markov chain monte carlo simulations. \emph{arXiv preprint
arXiv:2008.09589}.

\hypertarget{ref-ShahmoradiADS:2020}{}
Shahmoradi, A., \& Bagheri, F. (2020a). ParaDRAM: A Cross-Language
Toolbox for Parallel High-Performance Delayed-Rejection Adaptive
Metropolis Markov Chain Monte Carlo Simulations. \emph{arXiv e-prints},
arXiv:2008.09589.

\hypertarget{ref-ShahmoradiASCL:2020}{}
Shahmoradi, A., \& Bagheri, F. (2020b, August). ParaMonte: Parallel
Monte Carlo library.

\hypertarget{ref-Shahmoradi:2014}{}
Shahmoradi, A., \& Nemiroff, R. (2014). Classification and energetics of
cosmological gamma-ray bursts. In \emph{American astronomical society
meeting abstracts\# 223} (Vol. 223).

\hypertarget{ref-Shahmoradi:2015}{}
Shahmoradi, A., \& Nemiroff, R. J. (2015). Short versus long gamma-ray
bursts: A comprehensive study of energetics and prompt gamma-ray
correlations. \emph{Monthly Notices of the Royal Astronomical Society},
\emph{451}(1), 126--143.

\hypertarget{ref-Shahmoradi:2019}{}
Shahmoradi, A., \& Nemiroff, R. J. (2019). A Catalog of Redshift
Estimates for 1366 BATSE Long-Duration Gamma-Ray Bursts: Evidence for
Strong Selection Effects on the Phenomenological Prompt Gamma-Ray
Correlations. \emph{arXiv e-prints}, arXiv:1903.06989.

\end{document}